\documentclass[aps,pra,reprint,showpacs,superscriptaddress,twocolum]{revtex4-1}

\usepackage{amsmath}
\usepackage{amssymb}
\usepackage{graphicx}
\usepackage{hyperref}
\usepackage{epstopdf}
\hypersetup{
    colorlinks=true,linkcolor=blue,citecolor=blue,
    filecolor=blue,urlcolor=blue,breaklinks=true
}

\RequirePackage{color}

\begin{document}

\title{Quantum state transfer in optomechanical arrays}

\author{G. D. de Moraes Neto}
\email{gdmneto@gmail.com}
\affiliation{
  Department of Physics and Astronomy,
  University College London,
  London WC1E 6BT, United Kingdom
}

\author{F. M. Andrade}
\email{fmandrade@uepg.br}
\affiliation{
  Department of Physics and Astronomy,
  University College London,
  London WC1E 6BT, United Kingdom
}
\affiliation{
  Department of Computer Science,
  University College London,
  London WC1E 6BT, United Kingdom
}
\affiliation{
  Departamento de Matem\'{a}tica e Estat\'{i}stica,
  Universidade Estadual de Ponta Grossa,
  84030-900 Ponta Grossa-PR, Brazil
}

\author{V. Montenegro}
\email{v.montenegro.11@ucl.ac.uk}
\affiliation{
  Department of Physics and Astronomy,
  University College London,
  London WC1E 6BT, United Kingdom
}

\author{S. Bose}
\email{s.bose@ucl.ac.uk}
\affiliation{
  Department of Physics and Astronomy,
  University College London,
  London WC1E 6BT, United Kingdom
}

\date{\today}

\begin{abstract}
Quantum state transfer between distant nodes is at the heart of quantum
processing and quantum networking. Stimulated by this, we propose a
scheme where one can highly achieve quantum state transfer between sites
in a cavity quantum optomechanical network. 
There, each individual cell site is composed of a localized mechanical
mode which interacts with a laser-driven cavity mode via radiation
pressure, and photons exchange between neighboring sites is allowed. 
After the diagonalization of the Hamiltonian of each cell, we show that the
system can be reduced to an effective Hamiltonian of two decoupled
bosonic chains, and therefore we can apply the well-known results
regarding quantum state transfer in conjuction with an additional condition
on the transfer times. 
In fact, we show that our transfer protocol works for any arbitrary
quantum state, a result that we will illustrate within the red sideband regime. 
Finally, in order to give a more realistic scenario we take into account
the effects of independent thermal reservoirs for each site. Thus, solving the standard master equation within the Born-Markov
approximation, we reassure both the effective model as well as the
feasibility of our protocol.
\end{abstract}


\maketitle

\section{Introduction}
For quantum information processing purposes one often needs to transfer
a quantum state from one site to another \cite{Book.2010.Nielsen}, this
corresponding to the central goal in quantum networking schemes.
A wide range of physical systems able to carry information are used for
this end. 
For instance, proposals for quantum logical processing using trapped
atoms, for example, making use of traveling photons to transfer states
in cavity quantum electrodynamics (QED) \cite{PS.1998.1998.127} and
phonons in ion traps \cite{Nature.2000.403.269}. 

Although photons and phonons are individual quantum carriers on
themselves, several promising technologies for the implementation of
quantum information processing rely on collective phenomena to transfer
quantum states, such as optical lattices \cite{Nature.2003.425.937} and
arrays of quantum dots \cite{Nature.1998.393.133,*PRA.1998.57.120} just
to name a few.
It is therefore, a main goal to find physical systems that provide
robust quantum data bus (QDB) linking different quantum processors. 

In recent years, extensive theoretical research have been carried out on
the topic of state transfer in quantum networks, and many of them have
been conducted in several different systems and architectures
\cite{Book.2013.Nikolopoulos}. 

Interestingly, a plethora of results have been obtained based on
qubit-state transfer through spin chains considering different types of
neighbor (site-site) couplings
\cite{PRL.2004.92.187902,PRA.2006.73.032306}, as well as errors and
detrimental effects arising from network imperfections/non-idealities
\cite{PRL.1998.81.5932,NJP.2005.7.135,NJP.2007.9.79}. 

On the other hand, optical lattices constitute a promising platform for
quantum information processing, where both the coherent transport of
atomic wave packets \cite{PRL.2002.88.170408} as well as the evolution
of macroscopically entangled states \cite{PRA.2003.68.033609} have been
achieved. 

Furthermore, significant advances have been made in engineered (passive)
quantum networks, where the adjustment of static parameters leads to
quantum information tasks, such as, entanglement generation and state
transfer \cite{NJP.2004.6.36,*PRA.2005.71.032310,*PRA.2007.75.042319}. 

Motivated for all these aforementioned quantum systems towards quantum
networking/processing, we present the state transfer of quantum
information in optomechanical cavity systems 
---\textit{a promising growing field, where ``weak'' light-matter
interactions (trilinear radiation pressure interaction) take place
leading to interesting  quantum effects} \cite{RMP.2014.86.1391}. 

Specifically, we show that information encoded on polariton states,
\textit{i.e.}, photonic-phononic combined excitations, can be used to
transfer information from one site to another.
Additionally, the use of polariton states allow us to link both the
degrees of freedom of the quantized electromagnetic radiation field as
well as the mechanical mode.
Furthermore, polaritons permit undemanding manipulations with an
external laser field.
In fact, quantum state transfer of polaritonic qubits (photonic-atomic
excitations) in a coupled cavity system have been demonstrated
\cite{JMO.2007.54.2307,*PRA.2011.84.032339}. 

We would like to stress that, recent works on network of coupled
optomechanic cells \cite{Nature.2009.462.78} and light storage
\cite{NJP.2011.13.23003} have been introduced.
Also, collective effects as synchronization\cite{PRL.2011.107.043603}
quantum phase transitions \cite{PRA.2012.86.033821} and generation of
entanglement \cite{PRA.2012.86.042306} have been proposed in the
optomechanical field. 

Moreover, in earlier studies of quantum state transfer in optomechanical
systems relies on some sort of external control in the realm of active
small networks \cite{PRL.2012.108.153603,PRA.2015.91.032309} or quantum
state transfer only between mechanical modes \cite{NJP.2012.14.125005}.
The most straightforward approach in this context pertains to a sequence
of SWAP gates, which ensure the successive transfer of the state between
neighboring sites.
While intuitively simple, active networks are considered to be very
susceptible to errors ---\textit{which they are accumulated in each
operation applied during the transfer}, as well as to dissipation and
detrimental effects due to decoherence \cite{Book.2013.Nikolopoulos}. 

However, alternative strategies are based on the idea of eigenmode
mediated state transfer and rely on a perturbative coupling and ensure
resonance between the common frequency of the sender and the receiver
and a single normal mode of the QDB
\cite{NJP.2005.7.73,PRL.2011.106.040505} or a tunneling-like mechanism,
described by a two-body Hamiltonian, which allows either a bosonic or a
fermionic state to be transferred directly from the sender to the
receiver, without populating the QDB \cite{PRA.2012.85.052303}. 

In this manuscript, we envisage the quantum state transfer from a sender
to a receiver in an array of optomechanical cells.
There, each cell is composed of a localized mechanical mode that
interacts with a laser-driven cavity mode via radiation pressure, and
therefore photons can hop between neighboring sites. 

In addition, we show how to design the parameters that allow us perfect
state transfer of an arbitrary quantum state. In fact, two-way
simultaneous communications for different pairs of sites without mutual
interference. We stress that the linearization of the non-linear
optomechanical Hamiltonian does not constitute a major restriction. For
example, for driven optomechanical systems in the strong single-photon
regime, we can both transfer information encoded in polariton states
arising from ion trap-like Hamiltonian \cite{PRA.2013.88.063819} as well
as dark states in optomechanical systems \cite{PRA.2013.87.053849}. 

Finally, we illustrate the effectiveness of our protocol when each cell
is in contact with a thermal environment and under the red sideband
regime. 

\begin{figure*}
  \centering
  \textbf{(a)} \includegraphics[width=1.8\columnwidth]{fig1a}\\
  \vspace{1cm}
  \textbf{(b)} \includegraphics[width=0.8\columnwidth]{fig1b}
  \caption{(Color online)
     (a) Sketch of an array of $N$ optomechanical cells, each
     of these cells consists of a mechanical mode of frequency
     $\omega_{m}^{n}$ coupled via radiation pressure to a cavity mode of
     frequency $\omega_{r}^{n}$. 
     The optical mode is driven by a laser at frequency
     $\omega_{p}^{n}$ and cells are coupled by evanescent coupling
     between nearest neighbors cavities with hoping strength $J_{n}$.
     (b) Schematic of the effective model, two decoupled
     bosonic chains with polaritonic energies $ \Omega_{-}^{n} $, 
     $\Omega_{+}^{n} $ and neighbor-site hopping $\lambda_{n}$ and
     $\zeta_{n}$ respectively.} 
  \label{fig:fig1}
\end{figure*}

\section{The Model}

We consider a one-dimensional array of $N$ optomechanical cells, each of
these cells consists of a mechanical mode of angular frequency
$\omega_{m}^{n}$ coupled via radiation pressure to a cavity mode of
angular frequency $\omega_{r}^{n}$.
In addition, we consider an external laser driving the optical mode at
angular frequency $\omega_{p}^{n}$, as schematically depicted in
Fig. \ref{fig:fig1}(a).

Following the standard linearization procedure for driving optical modes
in optomechanical cavities, we can recast the following Hamiltonian (in
units of Planck constant, \textit{i.e.}, $\hbar=1$) 
\begin{equation}
  \hat{H}_{n}^{L}=
  -\Delta_{p}^{n}\hat{a}_{n}^{\dagger}\hat{a}_{n}
  +\omega_{m}^{n}\hat{b}_{n}^{\dagger}\hat{b}_{n}
  -G_{n}(\hat{b}_{n}+\hat{b}_{n}^{\dagger})
  (\hat{a}_{n}+\hat{a}_{n}^{\dagger}),
  \label{cell}
\end{equation}

where, the mechanical (optical) mode of the $n$-th cell is associated
with the bosonic operator $\hat{b}_{n}$ ($\hat{a}_{n}$);
$\Delta_{p}^{n}=\omega_{p}^{n}-\omega_{r}^{n}$ is the pump detuning from
cavity resonance, $g_{n}$ corresponds to the single-photon coupling rate
and $G_{n}=\alpha_{n}g_{n}$ is the effective optomechanical coupling
strength proportional to the laser amplitude. 

Here the cells are coupled by evanescent coupling between nearest
neighbors cavities with hoping strength $J_{n},$ an interaction
described by 
\begin{equation}
  \hat{H}_{I}=
  \sum_{n=1}^{N-1}J_{n}
  (\hat{a}_{n}^{\dagger}\hat{a}_{n+1}+\hat{a}_{n+1}^{\dagger}\hat{a}_{n}).
\end{equation}

As seen from the above Hamiltonian $\hat{H}_{n}^{L}$ (with
$\Delta_{p}^{n}<0$), we can readily notice two linearly coupled quantum
harmonic oscillators.
To obtain the relevant decoupled effective Hamiltonian, we proceed to
diagonalization of the Hamiltonian using the usual Bogoliubov transformation
as following:  
\begin{align}
  \hat{A}_{n} = {}
  &
    \mathcal{N}_{-}
    \left[
    \Delta_{1}^{n}\left(\Omega_{-}^{n}\right)\hat{a}_{n}^{\dag}
   +\Delta_{2}^{n}\left(\Omega_{-}^{n}\right)\hat{b}_{n}^{\dag}
    \right.
   \nonumber \\
  &
  \left.
   +\Delta_{3}^{n}\left(\Omega_{-}^{n}\right)\hat{a}_{n}
   +\Delta_{4}^{n}\left(\Omega_{-}^{n}\right)\hat{b}_{n}
    \right],
    \nonumber\\
  \hat{B}_{n}  = {}
  &
    \mathcal{N}_{+}
    \left[
    \Delta_{1}^{n}\left(\Omega_{+}^{n}\right)\hat{a}_{n}^{\dag}
   +\Delta_{2}^{n}\left(\Omega_{+}^{n}\right)\hat{b}_{n}^{\dag}
    \right.
    \nonumber \\
  &
    \left.
   +\Delta_{3}^{n}\left(\Omega_{+}^{n}\right)\hat{a}_{n}
   +\Delta_{4}^{n}\left(\Omega_{+}^{n}\right)  \hat{b}_{n}
    \right],
\end{align}
with eigenvalues
\begin{align}
  (\Omega_{\mp}^{n})^{2}= {}
  &
    \frac{{\Delta_{p}^{n}}^{2}+{\omega_{m}^{n}}^{2}}{2}
    \nonumber \\
  &
    \mp
    \frac{1}{2}
    \sqrt{
    \left(
    (\Delta_{p}^{n})^{2}-(\omega_{m}^{n})^{2}
    \right)^{2}
    -16G_{n}^{2}\Delta_{p}^{n} \omega_{m}^{n}},
\end{align}
where we have defined
\begin{align*}
  \Delta_{1}\left(  \Omega_{\mp}^{n}\right)   = {}
  &
    2G_{n}^{2}\omega_{m}^{n}
    -\left(\Omega_{\mp}-\omega_{m}^{n}\right)
    \left(
    \Omega_{\mp}^{n}
    -\left\vert \Delta_{p}^{n}\right\vert
    \right)  
    \left(\Omega_{\mp}^{n}+\omega_{m}^{n}\right),\\
  \Delta_{2}\left(  \Omega_{\mp}^{n}\right)   = {}
  &
    G_{n}\left[  \left(
    \Omega_{\mp}-\left\vert \Delta_{p}^{n}\right\vert \right)  \left(  \Omega
    _{\mp}-\omega_{m}^{n}\right)  \right]  ,\\
  \Delta_{3}\left(  \Omega_{\mp}^{n}\right)   = {}
  &
    2G_{n}^{2}\omega_{m}^{n},\\
  \Delta_{4}\left(  \Omega_{\mp}^{n}\right)   = {}
  &
    G_{n}\left(  \Omega_{\mp
    }-\left\vert \Delta_{p}^{n}\right\vert \right)  \left(  \Omega_{\mp}%
+\omega_{m}^{n}\right),
\end{align*}

and normalization
\begin{align}
  \frac{1}{\mathcal{N}_{n,\mp}^{2}} = {}
  &
    \left[\Delta_{3}\left(\Omega_{\mp}^{n}\right)\right]^{2}
   +\left[\Delta_{4}\left(\Omega_{\mp}^{n}\right)\right]^{2}
    \nonumber \\
  &
    -\left[\Delta_{1}\left(\Omega_{\mp}^{n}\right)\right]^{2}
    -\left[\Delta_{2}\left(\Omega_{\mp}^{n}\right)\right]^{2}.
\end{align}

Therefore, the total Hamiltonian
$\hat{H}=\hat{H}_{n}^{L}+\hat{H}_{I\text{ }}$ in the polariton basis can
be rewrite as
\begin{align}
  \tilde{H} = {}
  &
    \sum_{n=1}^{N}
    \Omega_{^{-}}^{n}\hat{A}_{n}^{\dagger}\hat{A}_{n}
   +\Omega_{^{+}}^{n}\hat{B}_{n}^{\dagger}\hat{B}_{n}
    \nonumber \\
  &
   +\sum_{n=1}^{N-1}
    (\lambda_{n}\hat{A}_{n}^{\dagger}\hat{A}_{n+1}
    +\zeta_{n}\hat{B}_{n}^{\dagger}\hat{B}_{n+1}+\rm{H.c.}),
    \label{Hp}
\end{align}

with the effective tunneling strength
\begin{align*}
  \lambda_{n} = {}
  &
   J_{n}\mathcal{N}_{n,-}\mathcal{N}_{n+1,-}
    \left[
    \Delta_{1}\left(\Omega_{-}^{n}\right)
    \Delta_{1}\left(\Omega_{-}^{n+1}\right)
    \right.
    \\
  &
    \left.
    +\Delta_{3}\left(  \Omega_{-}^{n}\right)
    \Delta_{3}\left(  \Omega_{-}^{n+1}\right)
    \right]
\end{align*}
and
\begin{align*}
  \zeta_{n} = {}
  &
    J_{n}\mathcal{N}_{n,+}\mathcal{N}_{n+1,+}
    \left[
    \Delta_{2}\left(\Omega_{+}^{n}\right)
    \Delta_{2}\left(\Omega_{+}^{n+1}\right)
    \right.
  \\
  &
    \left.
    +\Delta_{4}\left(\Omega_{+}^{n}\right)
    \Delta_{4}\left(\Omega_{+}^{n+1}\right)
    \right].
\end{align*}

It is important to point out that in deriving the above expression,
terms like $A_{i}^{\dagger}A_{i+1}^{\dagger}$ and
$A_{i}^{\dagger}B_{i+1}$ have been neglected due to the usual
rotating-wave approximation (RWA), which remains valid for 
\begin{align*}
  \Omega_{\mp}^{n},
  \left\vert \Omega_{+}^{n}-\Omega_{-}^{n}\right\vert
  \gg
  \sqrt{
  \sum_{n}^{N}
  \left\langle\hat{a}_{n}^{\dagger}\hat{a}_{n}\right\rangle
 +\left\langle\hat{b}_{n}^{\dagger}\hat{b}_{n}\right\rangle
  }
  (\lambda_{n}+\zeta_{n}).
\end{align*}

Now, it is straightforward to observe under the above mapping that the
original full Hamiltonian of a unidimensional array of optomechanical
cells becomes equivalent to a Hamiltonian of two distinct bosonic
chains, this Hamiltonian being the central result of this manuscript.
Scenario schematically illustrated in Fig. \ref{fig:fig1}(b).
Because of the effective structure achieved above, \textit{i.e.}, two
independent chains, we are now in position to take advantage of the
well-known results on quantum state transfer. 

As known from any state transfer scheme, the set of couplings parameters
$\left\{\lambda_{n},\zeta_{n}\right\}$ as well as energies
$\Omega_{\mp}^{n}$ defines the transfer time $\tau.$
Here, we point out that our protocol requires that the transfer time for
both polaritons $\hat{A}^{\dagger}\hat{A}$ and
$\hat{B}^{\dagger}\hat{B}$ has to be the same or at least an odd
multiple of each other. 

To illustrate this point, we will consider the red-detuned regime
$\Delta_{p}^{n}\approx-\omega_{m}^{n}$, thus the Hamiltonian
(\ref{cell}) can be simplified as 
\begin{equation}
  \hat{H}_{n}^{\rm{red}}=
  \omega_{m}^{n}(\hat{a}_{n}^{\dagger}\hat{a}_{n}
  +\hat{b}_{n}^{\dagger}\hat{b}_{n})-G_{n}(\hat{b}_{n}\hat{a}_{n}^{\dagger}
  +\hat{b}_{n}^{\dagger}\hat{a}_{n}).
\end{equation}

To obtain the diagonal form of the above expression, we consider the
operators 
\begin{equation}
\hat{A}_{n}=\frac{(\hat{a}_{n}+\hat{b}_{n})}{\sqrt{2}},
\qquad
\hat{B}_{n}=\frac{(\hat{a}_{n}-\hat{b}_{n})}{\sqrt{2}}
\end{equation}

with eigenvalues $\omega_{A}^{n}=\omega_{m}^{n}-G_{n}$ and
$\omega_{B}^{n}=\omega_{m}^{n}+G_{n}$, respectively.

For the strongly off-resonant regime ($G_{n} \gg J_{n})$ together with
the RWA, we can recast the following polariton Hamiltonian 
\begin{align}
  \hat{H} = {}
  &
    \sum_{n=1}^{N}
    \left(
    \omega_{A}^{n}\hat{A}_{n}^{\dagger}\hat{A}_{n}
   +\omega_{B}^{n}\hat{B}_{n}^{\dagger}\hat{B}_{n}
    \right)
    \nonumber \\
  &
    +\sum_{i=1}^{N-1}\frac{J_{n}}{\sqrt{2}}
    \left(
    \hat{A}_{n}^{\dagger}\hat{A}_{n+1}
    +\hat{B}_{n}^{\dagger}\hat{B}_{n+1}+\rm{H.c.}
    \right).
\label{pol}
\end{align}

Now we proceed to choose a set of parameters that allows quantum state
transfer. For instance, a straightforward set can be found in
Ref. \cite{Book.2013.Nikolopoulos} corresponding to 
$\omega_{m}^{n} =\omega_{m}, G_{n} = G $ and 
$J_{n} = ({J}/{\sqrt{2}})\sqrt{n(N-n)}$,
which provides the same transfer time for each chain 
$\tau_{A}=\tau_{B} = {\pi}/{J}$. 

Therefore, regardless a relative phase depending on $\omega_{A}$ and
$\omega_{B}$ which is fixed and known, and hence, it can be amended, any
optomechanical state can be transferred only ensuring the $G \gg J$
regime together with $J_{n} = ({J}/{\sqrt{2}})\sqrt{n(N-n)}$. 

However, we stress that any other protocol could have been chosen for
this purpose. 
For example, schemes based on eigenmodes, where one of many
possibilities that permit quantum state transfer is the following set of
parameters: 
$J_{1}=J_{N-1}=\lambda \ll J_{k}=J \ll G,k=2\ldots N-2,N$ being an odd
number and $\omega_{m}^{n}=\omega_{m},G_{n}=G$.

On the other hand, on resonant schemes
\cite{NJP.2005.7.73,*PRL.2011.106.040505} the shorter transfer time 
possible corresponds to $\tau_{A}=\tau_{B}=(\pi/\lambda)\sqrt{2(N+1)}$,
and for tunneling-like protocol \cite{PRA.2012.85.052303}
with same parameters and conditions
$\omega_{m}^{1}=\omega_{m}^{N}=\omega_{m}+\delta$ and
$\lambda \ll \delta\ll J$, we obtain transfer times
$\tau_{A}=\tau_{B}={N\pi\delta}/{2\lambda^{2}}$.

Finally, it is worth stressing that, the effect of a phononic hop term
between neighboring sites only change the strength of $\lambda_{n}$ and 
$\zeta_{n}$.

\section{Dissipative mechanisms}

In this section, in a step towards a more realistic model we take into
account decoherence and dissipation.
To fulfill this goal, we employ the standard formalism for open quantum
systems, \textit{i.e.}, we solve the dynamics of the optomechanical
array using the master equation in Lindblad form within the Born-Markov
approximation. 

Furthermore, we numerically investigate the effectiveness of our model
computing the fidelity for the state transfer considering engineered hop
couplings between cells where each cell is considered in the
red-sideband regime. 

The master equation for the composite coupled system is given as
\begin{align}
  \frac{d\hat{\rho}}{dt} = {}
  &
    -i\left[\hat{H},\hat{\rho}\right]\\
  &  
    +\sum_{n=1}^{N}
    \frac{\kappa_{n}}{2}(1+\overline{n}_{c})
    \mathcal{D}\left[\hat{a}_{n}\right]\hat{\rho}
    +\frac{\kappa_{n}}{2}\overline{n}_{c}
    \mathcal{D}\left[\hat{a}_{n}^{\dagger}\right]\hat{\rho}
      \nonumber \\
  &
    +\frac{\gamma_{n}}{2}(1+\overline{n}_{m})\mathcal{D}
    \left[\hat{b}_{n}\right]  \hat{\rho}
    +\frac{\gamma_{n}}{2}\overline{n}_{m}\mathcal{D}
    \left[\hat{b}_{n}^{\dagger}\right]
    \hat{\rho},
    \label{master}
\end{align}
where $\hat{H}=\sum_{n=1}^{N}\hat{H}_{n}^{\rm{red}}+\hat{H}_{I}$ and the
Lindblad term
\begin{equation}
  \mathcal{D}\left[\hat{O}\right] =
  2\hat{O}\rho\hat{O}^{\dagger}-
  \rho\hat{O}^{\dagger}\hat{O}-\hat{O}^{\dagger}\hat{O}\rho
\end{equation}

takes into account the dissipative mechanisms of the optics (mechanics)
in contact with a thermal reservoir with occupation number
$\overline{n}_{c}$ $(\overline{n}_{m})$, where the photon (phonon) decay
rate is given by $\kappa_{n}(\gamma_{n})$. 

\begin{figure}
  \centering
  \includegraphics*[width=1\columnwidth]{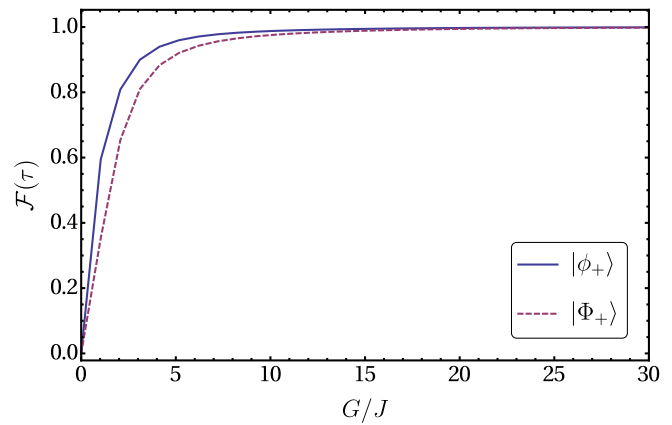}
  \caption{(Color online) 
    The figure shows the dynamics of the transfer fidelity at time 
    $\tau=\pi/J$ as a function of $G/J$.
    The states 
    $|\phi_{+}\rangle=({1}/{\sqrt{2}})\left(|1,0\rangle+|0,1\rangle\right)$
    and 
    $|\Phi_{+}\rangle=({1}/{\sqrt{2}})\left(|2,0\rangle+|0,2\rangle\right)$
    corresponds to the initial states of the sender. }
  \label{fig:fig2}
\end{figure}

\begin{figure}
  \centering
  \includegraphics*[width=1\columnwidth]{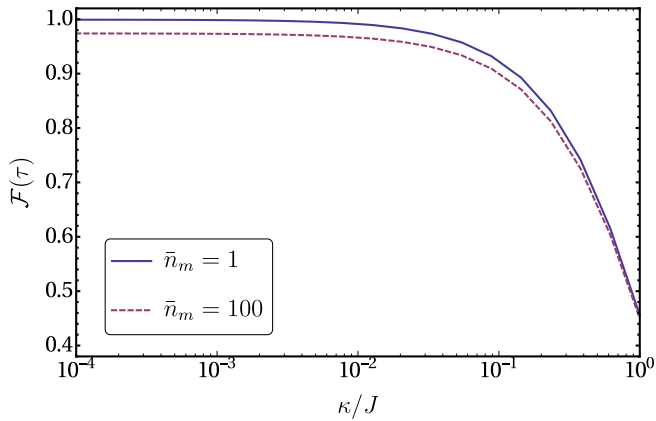}
  \caption{(Color online) 
    We illustrate the fidelity of the transfer process $(\tau=\pi/J)$ as
    a function of $\kappa/J$ (log-axis) for two mechanical phonon bath
    occupation number $\overline{n}_{m}=100$ and $\overline{n}_{m}=1$.
    To exhibit our findings, we consider the following feasible
    parameters in optomechanical crystals \cite{Nature.2011.478.89} in
    the microwave regime; 
    $\omega_{m}/2\pi=3.68\times10^{9},\gamma/2\pi=35\times10^{3}$, 
    $\overline{n}_{c}=0.005$, $G=({J}/{4})\times10^{2}=5\times10^{9}$.}
  \label{fig:fig3}
\end{figure}

Needless to say that the first non-trivial quantum network in passive
schemes is composed of four sites. Hence, for computational time
purposes, we will exemplify our findings considering an array of four
cells where the coupling fulfill $J_{n}=({J}/{\sqrt{2}})\sqrt{n(N-n)}$
and $\omega_{m}^{n}=\omega_{m}$.

To validate the polariton Hamiltonian (\ref{pol}), we present the closed
evolution of the transfer fidelity at time $\tau=\pi/J$ as a function of
$G/J$, see Fig. (\ref{fig:fig2}). 

In order to compute the fidelity of the transferred quantum state, we
solved the closed quantum system dynamics (running the simulation in
QuTiP \cite{CPC.2012.183.1760}) considering the sender initially in the
state 
$|\phi_{+}\rangle=({1}/{\sqrt{2}})\left(|1,0\rangle+|0,1\rangle\right)$
or 
$|\Phi_{+}\rangle=({1}/{\sqrt{2}})\left(|2,0\rangle+|0,2\rangle\right)$
(we have used the following notation 
$|a,b\rangle = |a\rangle_{optics}\otimes|b\rangle_{mechanics}$), where
all the other cells are in the vacuum state. 

Moreover, for our illustrative red sideband detuning regime 
$(-\Delta_{p}\approx\omega_{m}\gg\gamma,\kappa)$ a well-known stability
condition \cite{Incollection.2009.Genes} given by
$G<({1}/{2})\sqrt{\omega_{m}^{2}+{(\gamma^{2}+\kappa^{2})}/{4}}$ come
into sight, and therefore it must be observed throughout the quantum
state transfer protocol. 
On the other hand, in order to achieve a fidelity value close to the
unity, G has to be $G\approx({J}/{4})\times 10^{2}$ (as seen in
Fig. \ref{fig:fig2}). 
The effect of both the stability condition (being an upper bound for G),
as well as the effectiveness of the fidelity ($\mathcal{F(\tau)}
\rightarrow 1$), have as a result the limitation of the maximum coupling
strength $J_{n={N}/{2}}=N{J}/{4}$ and consequently the maximum number of
cells. 

In Fig. \ref{fig:fig3}, we compute the fidelity for the transfer of an
initial quantum state given by $|\phi_{+}\rangle$ as a function of
$\kappa/J$ for two different mechanical phonon bath occupation number
$\overline{n}_{m}=100$ and $\overline{n}_{m}=1$, where we have used the
following currently experimental parameters in optomechanical crystals
\cite{Nature.2011.478.89} in the GHz regime;
$\omega_{m}/2\pi=3.68\times10^{9}$, $\gamma/2\pi=35\times10^{3}$,
$\overline{n}_{c}=0.005$, $G=({J}/{4})\times10^{2}=5\times10^{9}$. 
The high fidelity shown in Fig. \ref{fig:fig3} up to $J=10\kappa$ is an
expected result, since $k_{B}T<\hbar\omega_{m}$, and the threshold for
coherent operations take place when
$\max(J_{n})=\max(\gamma_{n},\kappa_{n})$.
Thus, to achieve transfer fidelities close to unity for an array with
$N=100$ cells (with the same set of parameters considered above), we can
then estimate the cavity linewidth as $\kappa \sim 10^{5}$. 

Finally, we point out that the hopping coupling reported in
\cite{PRL.2011.107.043603} is in the range of THz. 
Hence, to achieve the inequality $J<G$ within the stability region, we
should engineered optomechanical arrays with larger lattice spacing
and/or mechanical modes with frequencies above THz, being this last a
challenging experimental scenario.

\section{Conclusion}

We have thus advanced a theoretical proposal for quantum state transfer
in optomechanical arrays. 
Our proposal relies on a general scheme illustrated by polariton
transformation of the linearized Hamiltonian (\ref{Hp}) that allow us to
obtain an effective Hamiltonian of two decoupled bosonic networks. 

The central result of the present manuscript is the derivation of the
polariton Hamiltonian (\ref{Hp}), where we can bring previous results
from quantum state transfer protocols in bosonic networks. 
Specifically, we can apply any type of quantum state transfer scheme
with an extra additional condition, namely, that the rate between the
transfer times of both decoupled polaritonic chains must be an odd
number. 
Furthermore, we analyze the effects of dissipation and a possible
experimental implementation of our proposal in the red-sideband regime
with experimental accessible parameters. 

It is also important to point out that 
---\textit{although not reported   explicitly in this work}--- the
linearization of the non-linear optomechanical Hamiltonian does not
constitute a major restriction.
For instance, for driven optomechanical systems in the strong
single-photon regime, we can both transfer information encoded in
polariton states arising from ion trap-like Hamiltonian
\cite{PRA.2013.88.063819} as well as  dark states in optomechanical
systems \cite{PRA.2013.87.053849}.  

Moreover, even though we used a one-dimensional array in this work, any
other topology might be consider, such as lattices (2D) or crystals (3D)
setups. 

In addition, other interesting aspects to study are the ``pretty good
state transfer'' schemes in Ref. \cite{Book.2013.Nikolopoulos}, and the
generation of long distance quantum entanglement between sites
\cite{PRA.2012.85.052303}.

\section*{Acknowledgments}
This work was supported by the Conselho Nacional de Desenvolvimento
Cient\'{\i}fico e Tecnol\'{o}gico - CNPq 
(Grants No 203097/2014--9 (PDE), 460404/2014--8 (Universal) and
206224/2014--1 (PDE)).

\bibliographystyle{apsrev4-1}
\bibliography{qst}

\end{document}